\documentclass[aps,floatfix,notitlepage,prl,reprint,superscriptaddress,twocolumn]{revtex4-1}
\usepackage{graphicx,amsmath,bm,natbib,color}

\begin{document}

\title{Some assembly required: assembly bias in massive dark matter halos} 
\author{Chun Yin Ricky Chue}
\affiliation{Department of Astronomy, University of Illinois at Urbana-Champaign}
\author{Neal Dalal}
\affiliation{Perimeter Institute for Theoretical Physics}
\author{Martin White}
\affiliation{Department of Physics, University of California, Berkeley}
\affiliation{Department of Astronomy, University of California, Berkeley}

\begin{abstract}
We study halo assembly bias for cluster-sized halos.  
Previous work has found little evidence for correlations between large-scale bias and halo mass assembly history for simulated cluster-sized halos, in contrast to the significant correlation found between bias and concentration for halos of this mass.  This difference in behavior is surprising, given that both concentration and assembly history are closely related to the same properties of the linear-density peaks that collapse to form halos.  Using publicly available simulations, we show that significant assembly bias is indeed found in the most massive halos with $M\sim 10^{15}M_\odot$, using essentially any definition of halo age.  For lower halo masses $M\sim 10^{14}M_\odot$, no correlation is found between bias and the commonly used age indicator $a_{0.5}$, the half-mass time.  We show that this is a mere accident, and that significant assembly bias exists for  other definitions of halo age, including those based on the time when the halo progenitor acquires some fraction $f$ of the ultimate mass at $z=0$.  For halos with $M_{\rm vir}\sim 10^{14}M_\odot$, the sense of assembly bias changes sign at $f=0.5$.  We explore the origin of this behavior, and argue that it arises because standard definitions of halo mass in halo finders do not correspond to the collapsed, virialized mass that appears in the spherical collapse model used to predict large-scale clustering.  Because bias depends strongly on halo mass, these errors in mass definition can masquerade as or even obscure the assembly bias that is physically present.  More physically motivated halo definitions using splashback should be free of this particular defect of standard halo finders.  
\end{abstract}
\maketitle

The clustering of tracers of cosmological large-scale structure, such as galaxies, quasars, clusters, or voids, may be used to probe the clustering of the underlying matter field.  The clustering strength of any particular tracer does not exactly match the clustering of total matter, but instead is generally biased relative to matter clustering \citep{Kaiser1984}.  On large scales, in the linear regime of structure formation for standard cosmologies with cold dark matter and gravity described by Einstein's general relativity, the bias for any tracer tends towards a constant value that becomes independent of scale \citep[e.g.][]{BBKS,Scherrer1998}. 
For dark matter halos, the linear bias is a strong function of halo mass, with the most massive halos clustering far more strongly than typical dark matter particles, while the smallest halos cluster less strongly than typical particles \citep{Kaiser1984,BBKS,Cole1989}.  Qualitatively, one may think of highly biased halos ($b\gg1$)  as preferentially forming in regions of high density, while halos with low bias (e.g., $b<1$) tend to avoid high-density regions.

In addition to its mass dependence, halo bias can also depend on other halo properties such as mass assembly history \citep{Gao2005} or  properties like concentration, spin, etc.\ \citep{Gao2007}.  Although not as strong as the mass dependence, these secondary dependencies of halo bias can be quite significant, in some cases leading to variations in linear bias of more than a factor of 2 for halos of fixed mass.  Because secondary bias can be quite significant, a number of studies have explored the impact of such biases on the galaxy-halo connection; see Ref.\ \cite{Wechsler2018} for a recent review of this topic and for a more comprehensive review of work on secondary biases.  The most well-studied of these secondary biases have been assembly bias, the dependence of bias on mass assembly history (MAH), and the concentration bias, referring to the dependence on halo concentration.  In general, secondary biases exhibit significant mass dependence.  For example, the concentration bias actually reverses in sign as halo mass is varied, with high concentration associated with high bias for small halos but with low bias for the largest halos \citep{Wechsler2006}.

Much of this behavior in halo bias is not difficult to understand in the context of hierarchical structure formation \citep{Dalal2008}.  Because halos tend to arise from peaks of the linear density field \citep{BBKS,Ludlow2011}, the properties of halos are related to the properties of the corresponding initial peaks.  For example, peaks with steep slopes tend to produce halos with high concentration, while peaks with shallow slopes tend to lead to halos with low concentration \citep[e.g.][]{Dalal2010}.  Additionally, because the linear density field is continuous, the slopes of initial peaks are also correlated with their local environments.  At fixed peak height, peaks with steep slopes tend to be found in relatively lower density environments than peaks with shallow slopes.
This accounts for the concentration bias seen at high halo masses \citep{Dalal2008}, however this does not explain the opposite behavior seen at low halo mass.  At lower masses, another process starts to dominate over the effect of peak slopes in producing concentration bias (and assembly bias).  Among low-mass halos below the nonlinear mass scale  ($M\ll M_\star$), a significant fraction of order 20\% ceases to grow in mass, due to environmental effects.  Because halo concentration is related to assembly history \citep{Wechsler2002}, the halos that stop growing exhibit the highest concentrations.  At the same time, the environmental effects that shut down halo growth (e.g., strong tides or high velocity dispersion) are also associated with high density regions.  For this reason, at low masses high concentrations become correlated with high local density, i.e.\ high bias.  This effect is unimportant at the very highest masses because the biggest halos dominate their environments.  

A corollary of the argument explaining concentration bias is that very similar behavior should be found in assembly bias.  At high masses, the same peak properties that determine halo concentration also determine halo assembly histories, and at low masses, the environmental effects that lead to high concentration also arrest the growth of halo mass.  The expected assembly bias is indeed found in low-mass halos \citep{Gao2005}, but at higher masses, the evidence is far less clear. \citet{Gao2007} found no significant assembly bias at high mass in their simulations, and \citet{Mao2018} argued that cluster-sized halos exhibit no detectable assembly bias in $\Lambda$CDM simulations.  If correct, this result would be remarkable and would require a dramatic rethinking of halo formation in general.  The prediction of assembly bias follows from the continuity of the linear density field, given the known result that the formation of the most massive halos closely follows the prediction of the spherical collapse model \citep{Gunn1972} that formation occurs when the smoothed linear density reaches a critical value, $\bar\delta = \delta_{\rm c} \approx 1.686$ \citep{Dalal2008,Robertson2009}.  Since the linear density field is indeed continuous, the prediction of nonzero assembly bias at high mass would seem to be inescapable.  

Motivated by this surprising claim, we investigate halo assembly bias for massive cluster-sized halos in $\Lambda$CDM simulations.  Since we focus on only the most massive halos which tend to be rare, we utilize simulations with large volume.  Most of the results we present below are derived from the BigMDPL simulation \citep{Prada2012}, publicly available at \texttt{https://www.cosmosim.org} \citep{Riebe2013}.  This simulation is part of the MultiDark simulation suite, and contains $3840^3$ particles in a box of comoving side length of 2.5 $h^{-1}$ Gpc for a flat $\Lambda$CDM cosmology with $\Omega_m\approx 0.307$, $h=0.6777$, $\sigma_8=0.8228$ and $n_s=0.96$, corresponding to particle mass $m_p=2.36\times 10^{10} h^{-1} M_\odot$.  We use the Rockstar \citep{Rockstar} halo catalogs and merger trees publicly provided at \texttt{https://www.cosmosim.org}.  To derive mass accretion histories, we follow the main branch of the Rockstar merger tree, using the {\tt mmp} (most massive progenitor) flag.  As a sanity check, we have also examined other simulations, including the MDPL2 simulation from the same MultiDark suite, as well as a series of $L=640\, h^{-1}$ Mpc simulations run for this investigation.  As a check on the Rockstar results, we have computed halo catalogs and merger trees using a different method for the 640 Mpc boxes, as described in Ref.~\cite{Cohn2008}.  In all cases, we find results consistent with the BigMDPL simulation results, so the discussion below will focus on that simulation since it provides the best statistics due to its large volume.

For the BigMDPL simulation, we measure the linear bias for halo samples by first computing the halo-matter cross spectrum $P_c(k)$ and the matter auto-spectrum $P_m(k)$, and then defining the bias $b$ by a least-squares fit for $P_c(k) = b\,P_m(k)$ for $k<0.1\, h$ Mpc$^{-1}$.  Because the matter field is not made publicly available for this simulation, as a proxy for the matter field we use the set of all halos and subhalos with $M_{\rm peak}\geq 5\times 10^{11} h^{-1} M_\odot$ in the $z=0$ Rockstar catalog.  These halos should be nearly unbiased on large scales, but it is worth noting that formally all of our quoted bias values really correspond to the ratio $b/b_{\rm tracer}$ where $b_{\rm tracer}$ is the mean bias of our tracer population of subhalos.

To start, we first examine halos with $M_{\rm vir}=0.7-1\times 10^{15} h^{-1} M_\odot$.  Previous work has shown significant concentration bias for halos in this mass range, and the BigMDPL simulation gives consistent results.  Rank ordering the halos based on the concentration values reported in the Rockstar catalogs, we measure mean linear biases for the subsets with the highest 25\% and lowest 25\% of $c_{\rm vir}$.  The quartile with highest $c_{\rm vir}$ gives $b_{c-{\rm high}} = 4.4 \pm 0.08$, while the quartile with lowest concentration gives $b_{c-{\rm low}} = 5.2 \pm 0.08$, as expected for the concentration bias at these high halo masses.  

Next, we turn to assembly bias.  Similar to the concentration split, we can split halos into the oldest and youngest quartiles, using some definition of halo age.  In previous literature \citep{Gao2007,Mao2018}, the half-mass time $a_{0.5}$ has been the most common definition of age.  This is defined as the scale factor when a given halo's most massive progenitor first acquires a fraction $f=0.5$ of the final mass at $z=0$.  From the Rockstar merger trees, we can readily determine $a_f$ for any fraction including $f=0.5$.  Halos with a small $a_f$ assembled fraction $f$ of their mass relatively earlier, and therefore may be considered to be older, while conversely halos with larger $a_f$ may be considered to be younger.  If we split halos in this mass range ($M_{\rm vir}=0.7-1\times 10^{15} h^{-1} M_\odot$) then the top and bottom quartiles give $b_{a-{\rm high}} = 4.9 \pm 0.07$ and $b_{a-{\rm low}} = 4.6 \pm 0.07$.  Therefore we do find significant assembly bias in high mass halos, with the expected sign, but the amplitude is about half as strong as the concentration bias for the same halos.  We find similar results for even higher masses or from other simulations, albeit with larger uncertainties.  It is reassuring that this basic prediction of Gaussian statistics is confirmed, but the weaker amplitude relative to concentration bias is somewhat surprising.  One possibility is that $a_{0.5}$ may simply be noisier than concentration.  This quantity is derived by tracking $M_{\rm vir}$ along the merger tree, but $M_{\rm vir}$ itself is a noisy estimate of the true virialized mass in a halo for a variety of reasons, including the presence of substructure, or the fact that the nominal virial radius $r_{\rm vir}$ can be either larger or smaller than the actual virialized region around a halo, the splashback radius \citep{Diemer2014,Adhikari2014}.  

If the assembly bias seen using $a_{0.5}$ is weak simply due to noise in the MAH, then we could improve the significance by using the entire MAH to classify halos into `young' or `old'.  As is well known, halo mass accretion histories exhibit a variety of behaviors \citep[e.g.][]{Wechsler2002}, so there is little reason to expect an arbitrarily chosen number like $a_{0.5}$ to capture the aspects of halo assembly that relate to large-scale environment.  However, since the entire MAH has many degrees of freedom, it may not be immediately obvious what definition of age that we should use instead of $a_{0.5}$.

The approach that we use is to perform a linear operation on the MAH to assign a single number to each halo, and then rank order based on that number.  To choose what linear operation to perform on the MAH, note that we can predict how the MAH should change when we raise or lower the large-scale linear density, using Gaussian statistics and the spherical collapse model.  The starting point is again the spherical collapse result that collapse occurs when the linear density smoothed over radius $R$ exceeds the collapse threshold, $\bar\delta(R) \geq \delta_{\rm c}$.  The model predicts that the set of halos of mass $M$ therefore should have $\bar\delta(R_{\rm L}) = \delta_{\rm c}$, where $R_{\rm L}=(3M/4\pi\bar\rho_m)^{1/3}$ is the Lagrangian radius corresponding to mass $M$ in the notation of \citep{Dalal2010}.  The linear density profile interior to $R_{\rm L}$ determines the assembly history of that halo \citep{Dalal2008}.  Therefore, to predict how the assembly history changes when we vary the large-scale environment, we simply need to know the expected value of $\bar\delta(R)$ at $R<R_{\rm L}$ as a function of the large-scale environmental density $\delta_{\rm long}$.  This is readily determined from the Gaussian statistics of the linear density field.  In general, for Gaussian distributed quantities $\bm{X}$ and $\bm{Y}$ with zero mean, the expected value of $\bm{X}$ conditioning on the value of $\bm{Y}$ is given by 
\begin{equation}
\langle\bm{X}|\bm{Y}\rangle = \langle\bm{X}\bm{Y}\rangle
\langle\bm{Y}\bm{Y}\rangle^{-1}\bm{Y}.
\label{cond}
\end{equation}
In our case, $\bm{X}$ consists of the interior profile $\bar\delta(R)$ for $R<R_{\rm L}$, and $\bm{Y}$ consists of the pair of quantities $\bar\delta(R_{\rm L})=\delta_{\rm c}$ and $\delta_{\rm long}$ on some large scale.  For concreteness, we define $\delta_{\rm long}$ as the linear density smoothed with a top hat filter of radius 30 $h^{-1}$ Mpc. 

Eq.\ \eqref{cond} gives us the expected profile for a peak of size $R_{\rm L}$ in a background overdensity $\delta_{\rm long}$, and if we know the linear growth factor $D(a)$ as a function of $a$, we can translate that peak profile into a mass assembly history by setting the collapse radius at each time $a$ such that $\bar\delta(R)D(a) = \delta_{\rm c}$.  Since Eq.\ \eqref{cond} is linear in $\delta_{\rm long}$, then for small $\delta_{\rm long}$ the response of the halo MAH is also linear in $\delta_{\rm long}$.   If we think of the MAH as a vector $\bm{h}$, then its expected linear response to $\delta_{\rm long}$ may be written as $\bm{h}={\rm const}+\bm{g}\,\delta_{\rm long}$, where the vector $\bm{g}$ encodes the linear response computed above.  This immediately suggests a sensible choice for the linear operation to perform on the actual MAH to assign an age to each halo: the inner product between $\bm{h}$ and the expected response vector $\bm{g}$.  To define an inner product on the space of possible assembly histories, however, we need some notion of a metric on that space, i.e.\ a matrix to allow us to compute distances and dot products between vectors.  One obvious choice for this metric is the inverse covariance matrix of all MAH's for halos in the mass bin being considered, ${\bf C}_h^{-1} = (\langle\bm{h}\bm{h}\rangle-\langle\bm{h}\rangle\langle\bm{h}\rangle)^{-1}$.  

\begin{figure}
\centerline{\includegraphics[width=0.48\textwidth]{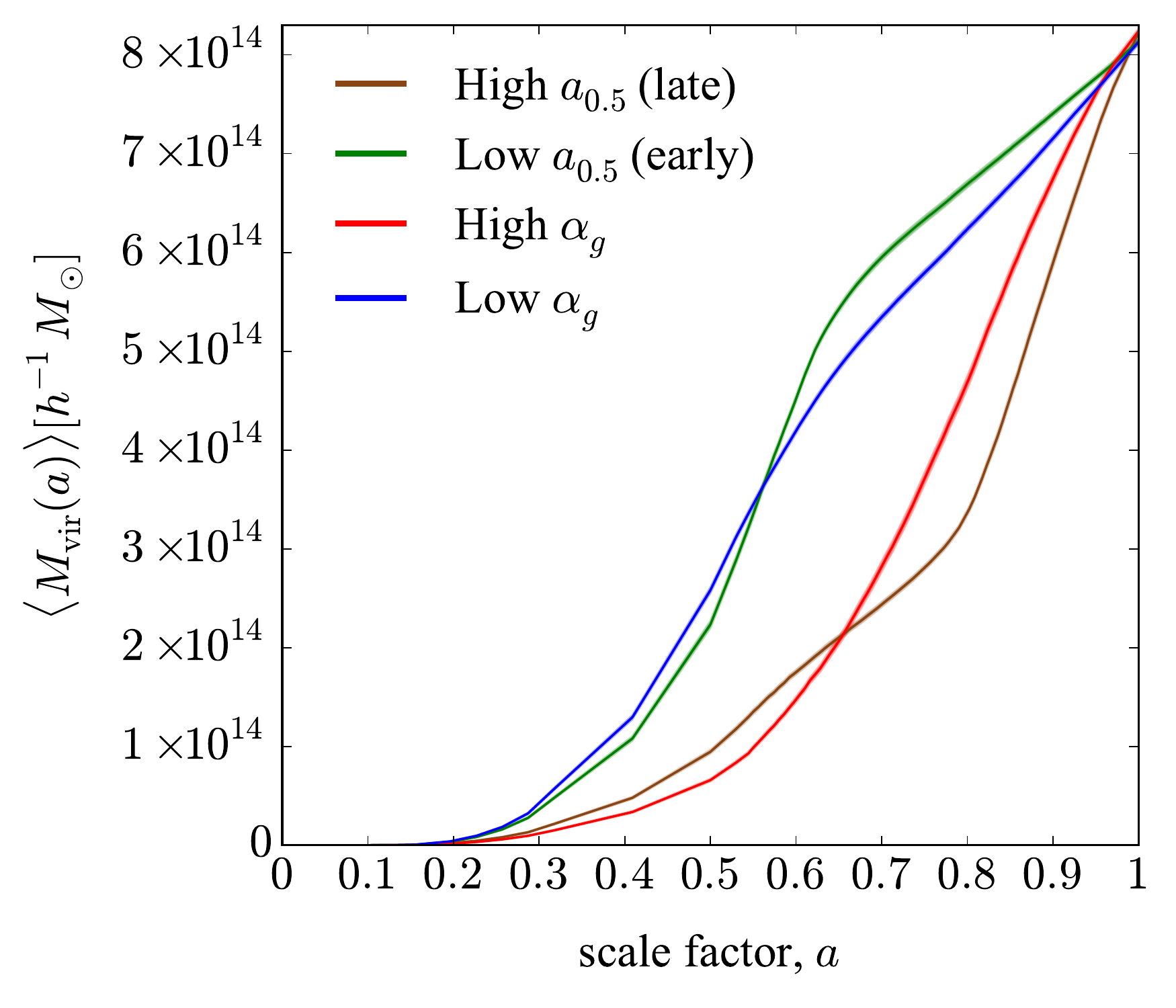}}
\caption{The different curves show stacked mass accretion histories for subsets of BigMDPL halos with $M_{\rm vir}=0.7-1\times 10^{15} h^{-1} M_\odot$.  The curves corresponds to the halos with the highest 25\% of $a_{0.5}$ (brown), the lowest 25\% of $a_{0.5}$ (green), the top 25\% of $\alpha_{\rm opt}$ (red), and bottom 25\% of $\alpha_{\rm opt}$ (blue).  The width of each curve corresponds to the $1-\sigma$ jackknife uncertainty on the mean MAH. As discussed in the text, $\alpha_{\rm opt}$ is a better indicator of large-scale bias than $a_{0.5}$, and it tends to split the halos more strongly on their early assembly histories ($f<0.5$) for this mass range.  
\label{bigmah}}
\end{figure}

Our procedure, therefore, is to define the `age' of each halo from its MAH $\bm{h}$ as
\begin{equation}
\alpha_g = \bm{g}^{\sf T}\cdot{\bf C}_h^{-1}\cdot\bm{h},
\end{equation}
where $\bm{g}$ is computed from Gaussian statistics as described above, and ${\bf C}_h^{-1}$ is computed from the ensemble of MAH's of the halo mass bin under consideration.  Defined in this way, halos with high $\alpha_g$ are expected to be more highly biased than halos with low $\alpha_g$, as long as halos are forming according to spherical collapse.  When we apply this age definition to halos in the same mass range ($M_{\rm vir}=0.7-1\times 10^{15} h^{-1} M_\odot$) considered above, the bias of the high $\alpha_g$ quartile is $b_{\alpha-{\rm high}} = 5.0\pm 0.07$, while the low $\alpha_g$ quartile gives $b_{\alpha-{\rm low}} = 4.5\pm 0.07$.  Evidently, using the entire MAH does enhance the amplitude of assembly bias, though the overall signal is still slightly smaller than the amplitude of the concentration bias.  In Fig.\ \ref{bigmah} we plot the stacked MAH's for the top and bottom quartiles of $\alpha_g$, along with stacked MAH's for the top and bottom quartiles of $a_{0.5}$.  The MAH's selected by $\alpha_g$ differ more at early times than the MAH's selected by $a_{0.5}$.

Therefore, the highest mass halos do exhibit clear assembly bias, as required theoretically.  This may seem to contradict previous results \citep{Gao2007,Mao2018}, but note that so far we have focused on halos with $M\sim 10^{15}M_\odot$, whereas previous works studied smaller clusters with $M\sim 10^{14}M_\odot$.  Therefore, we next consider halos with $M_{\rm vir} = 1-2\times 10^{14} h^{-1} M_\odot$.  When we split these halos using $a_{0.5}$, we do not find significant differences in the biases of the oldest or youngest halos.  This agrees with previous work, but is contrary to the results for the higher mass sample.  

\begin{figure}
\centerline{\includegraphics[width=0.48\textwidth]{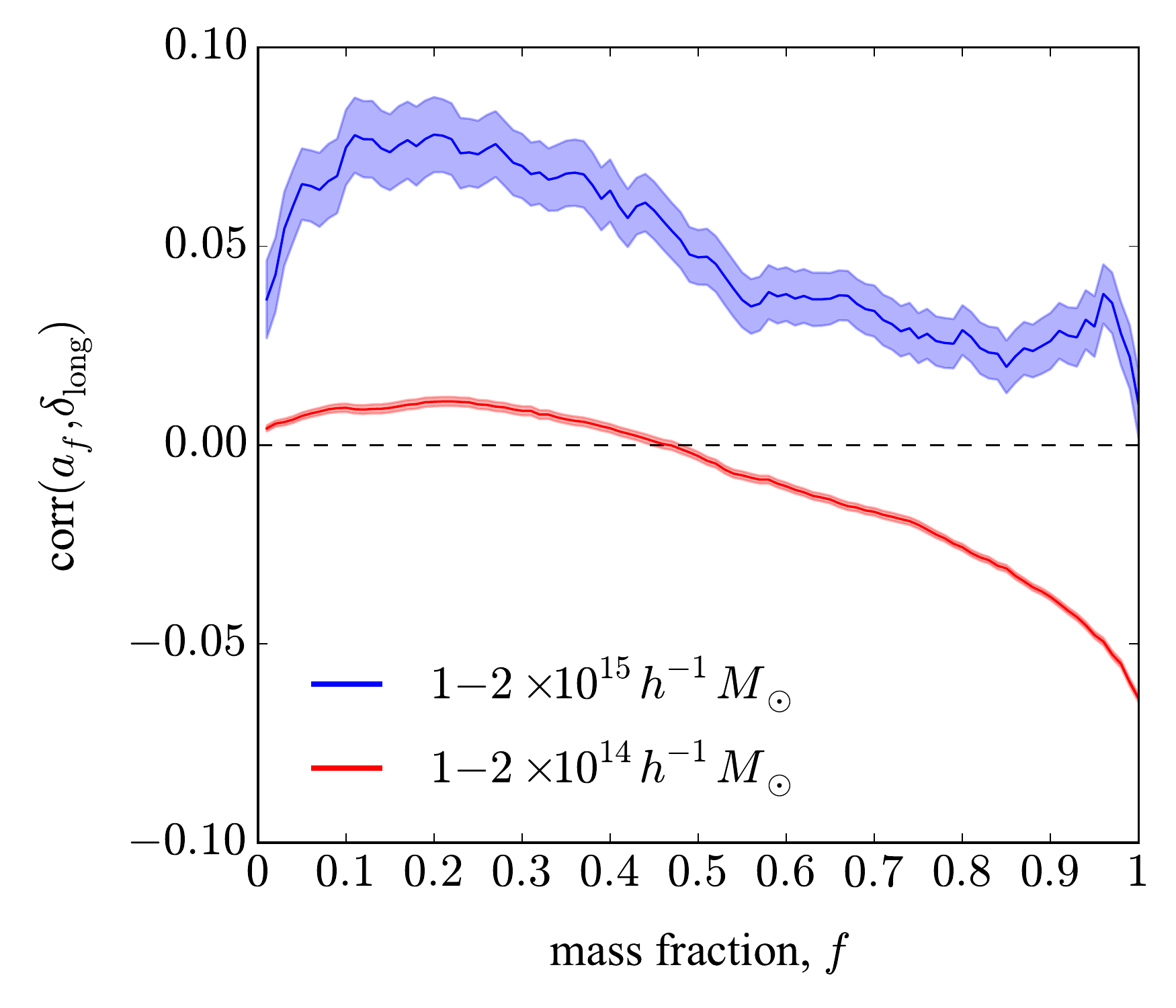}}
\caption{Plotted is the cross-correlation coefficient between the linear overdensity, smoothed on a scale of 30 $h^{-1}$ Mpc, and the time when the halo acquires fraction $f$ of its present-day mass, denoted $a_f$. The blue curve corresponds to high-mass halos in the mass range $M_{\rm vir}= 1-2\times 10^{15} h^{-1} M_\odot$, while the red curve is for low-mass halos with $M_{\rm vir}= 1-2\times 10^{14} h^{-1} M_\odot$.  The width of each band corresponds to the $1-\sigma$ uncertainty, determined by jackknife.
\label{ccc}}
\end{figure}

To understand this change in behavior, in Fig.\ \ref{ccc} we plot the cross-correlation coefficient between the large-scale density and the mass accretion history.  As above, the large-scale density is defined as $\delta_{\rm long}=\delta_{30}$, the overdensity smoothed over a scale of 30 $h^{-1}$ Mpc.  We characterize the MAH using $a_f$, the scale factor when a halo reaches fraction $f$ of its $z=0$ mass.  The cross-correlation coefficient is defined as 
\begin{equation}
{\rm{corr}}(a_f,\delta_{\rm{long}})=\frac{\langle (a_f-\bar a_f)(\delta_{\rm long}-\bar\delta_{\rm long})\rangle}{\left[\langle (a_f-\bar a_f)^2\rangle\langle(\delta_{\rm long}-\bar\delta_{\rm long})^2\rangle\right]^{1/2}},
\end{equation}
where $\bar a_f=\langle a_f\rangle$, $\bar\delta_{\rm long}=\langle\delta_{\rm long}\rangle$, and averages are computed over the sample of halos being considered.  
A positive cross-correlation means that increasing the large-scale density increases $a_f$, i.e.\ delays the time when the halo acquires mass fraction $f$.  

As Fig.\ \ref{ccc} shows, there is no significant correlation between large-scale density and $a_{0.5}$ for $M\sim 10^{14} M_\odot$ halos, but this appears to be an accident.  If we use some other fraction, like $a_{0.2}$ or $a_{0.8}$, then we do find significant correlations.  The early part of the MAH behaves similarly to the behavior for the $M\sim 10^{15} M_\odot$ halos: younger halos are associated with higher density.  But the later part of the MAH, for $f \lesssim 1$, has the opposite correlation.  The cross-over happens to occur near $f=0.5$, by accident.  Note that this is dependent on the mass of the sample.  For even lower masses, the cross-over occurs at even smaller $f$, and for higher masses it occurs at higher $f$ (or does not occur at all in the most massive halos, as shown in the blue curve in Fig.\ \ref{ccc}).  Note that the significant correlations between large-scale density and MAH that we find do not necessarily contradict the results of \citet{Mao2018}, who found that the stacked MAH's for halos with $M\sim 10^{14} M_\odot$ found in large-scale over-densities were very similar to those found in large-scale under-densities.  When we perform the same exercise, we also find similar MAH's with percent-level differences.  However, that is exactly the amplitude of difference that is expected.  The amplitude of density fluctuations on large scales in the linear regime is small by definition, percent-level for the scales of interest here.  Because the expected level of assembly bias is of order unity, not order 100, these percent-level overdensities on large scales should correspond to percent level variations in the MAH's, as observed.

The significant correlations at $f\neq 0.5$ imply that assembly bias is present in halos of this mass range, i.e.\ there are correlations between large-scale density and assembly history.  Accidentally, $a_{0.5}$ is insensitive to this assembly bias, however we can use other metrics for halo age to find significant assembly bias.  For example, we can once again use the `theoretical' template $\alpha_g$ to select old or young halos, which does indeed give nonzero assembly bias.  Alternatively, we can derive the optimal definition of halo age to maximize the difference in bias between old and young subsets.  We do so by cross-correlating halo MAH's with their large-scale density.  As before, we quantify the large-scale density as $\delta_{\rm long}=\delta_{30}$, the overdensity centered on a halo smoothed over a 30 $h^{-1}$ Mpc radius.  Similarly, again let us write $\bm{h}$ as the MAH for a halo.  If $\delta$ and $\bm{h}$ are Gaussian distributed then the optimal definition of age for a halo with a MAH $\bm{h}$ is given by $\alpha_{\rm opt} = \bm{d}^{\sf T}\cdot(\bm{h}-\bar{\bm{h}})$, where
\begin{equation}
\bm{d} = {\bf C}_h^{-1} \langle(\bm{h}-\bar{\bm{h}})(\delta_{\rm long}-\bar\delta_{\rm long})\rangle.
\label{opt}
\end{equation}
In other words, we take the inner product of each MAH with the part of the MAH correlated with large-scale density, where the inner product over the space of MAH's is defined using the inverse covariance of MAH's as the metric.  Note that to avoid over-fitting, when evaluating Eq.\ (\ref{opt}) for each cluster, we exclude all halos in the spatial octant centered on that cluster in computing the ensemble averages.  In labeling this definition optimal, what we mean is that this definition should maximize the difference in large-scale bias of the two samples, using only the mass accretion histories, as long as the underlying assumption of Gaussianity is approximately satisfied. Nongaussianity will make this definition sub-optimal for the purpose of splitting halos into high-bias and low-bias subsets, but as long as we do detect assembly bias any suboptimality does not impact our conclusions significantly.

\begin{figure}[tb]
\centerline{\includegraphics[width=0.48\textwidth]{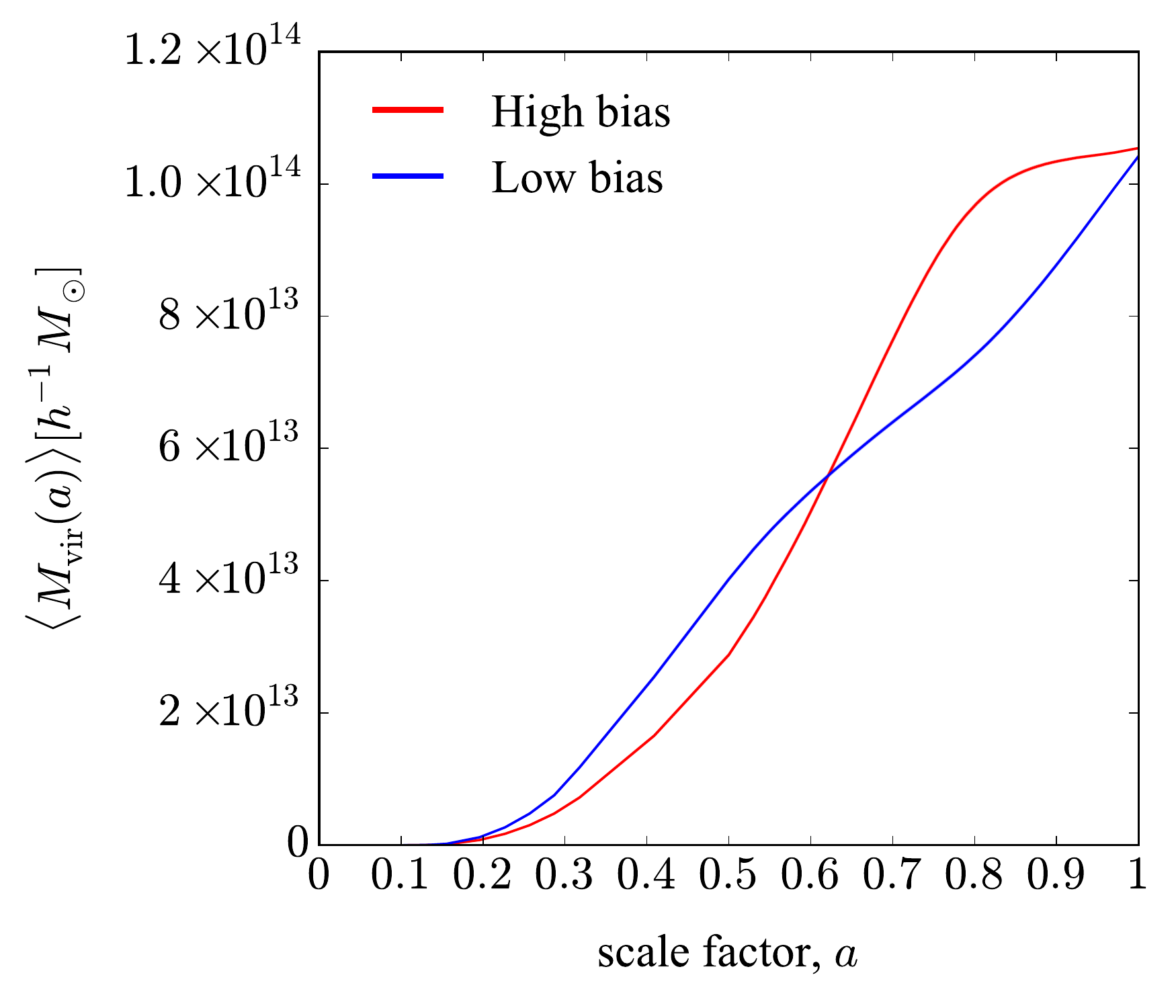}}
\caption{BigMDPL halos with $M_{\rm vir}=1-1.1\times 10^{14} h^{-1} M_\odot$.  Plotted are the stacked (average) MAHs for isolated halos in the top (red) and bottom (blue) quartiles of $\alpha_{\rm opt}$ as defined in Eq.\ (\ref{opt}).  The width of each curve corresponds to the $1-\sigma$ jackknife uncertainty on the mean MAH.  We have used a narrow mass bin in order to enforce that both subsets have the same average mass at $z=0$.  
\label{optmah}}
\end{figure}

Fig.\ \ref{optmah} shows the average MAH's for the halos in this mass range, split into top and bottom quartiles using $\alpha_{\rm opt}$.  The quartile with high $\alpha_{\rm opt}$ (red curve) has a mean linear bias $b= 2.2\pm 0.02$, while the quartile with low $\alpha_{\rm opt}$ (blue curve) has a mean linear bias $b=2.0 \pm 0.02$.  As expected, there is significant assembly bias among halos in this mass range, in that we can split halos into samples with higher or lower bias using only their MAH's.  It is difficult to say which subset is older or younger: at low mass fractions, the blue subset is significantly older, while at high mass fractions, the red subset is significantly older.  

One striking property of the red curve in Fig.\ \ref{optmah} is that the mean MAH nearly plateaus at late times, $a>0.85$.  This lack of growth in halo mass is quite surprising for cluster-sized halos.  Even if the physical mass distribution around the cluster remains static in time, the nominal virial mass will grow simply due to the decrease in the mean matter density as the universe expands, an effect called pseudo-evolution \citep{Diemer2013}.  For a static mass profile $M(r)$ around a halo, pseudo-evolution gives a minimum growth rate of $d\log M_{\rm vir}/d\log a = (d\log\rho_{\rm vir}/d\log a)\times \left[1 + 3/ (d\log\bar\rho/d\log r|_{r=r_{\rm vir}})\right]$, where $\bar\rho(r) = 3M(r)/(4\pi r^3)$, and $\rho_{\rm vir} = \Delta_{\rm vir}\rho_m$ for virial overdensity $\Delta_{\rm vir}$ \citep{Bryan1998} and mean matter density $\rho_m = \Omega_m\rho_{\rm crit}$.  Since these clusters tend to have somewhat low concentrations, e.g.\ $c_{\rm vir}\sim 6$, then for NFW outer profiles we would expect $d\log M_{\rm vir}/d\log a > 0.5 $ even if the density profiles around the halos remain static in time. Of course, the outer profiles of these halos can be steeper than NFW, due to the splashback feature \citep{Diemer2014}, but that steepening would only affect the pseudo-evolution rate of $M_{\rm vir}$ when the splashback radius is $r_{\rm sp} \lesssim r_{\rm vir}$, which only occurs for high accretion rates \citep{Adhikari2014}.  For the low growth rates shown in the red curve ($d\log M_{\rm vir}/d\log a \approx 0.18$), the splashback radius should be  outside $r_{\rm vir}$, implying that the NFW profile should be a reasonable approximation.  We will return to this topic later, but for now, the point is that  the observed growth rate in this subset of clusters is even less than the minimal pseudo-evolution rate for static mass distributions. 
In order for the average $M_{\rm vir}$ to grow so slowly with time, mass must be physically removed from within $r_{\rm vir}$ for at least some fraction of the clusters in the red subset.

One possible explanation for this could be that many of the halos in the red subset are in extreme environments capable of stripping mass from these cluster-sized halos.  To check for this, we search for more massive neighbors ($M_{\rm vir} \geq 2\times 10^{14} h^{-1} M_\odot$) within a few Mpc of these clusters.  
We find that only a tiny, percent-level fraction of the halos (excluding subhalos) have massive nearby neighbors capable of tidally stripping the clusters.

\begin{figure}
\centerline{\includegraphics[width=0.48\textwidth,trim=0 2.5in 0 2.5in]{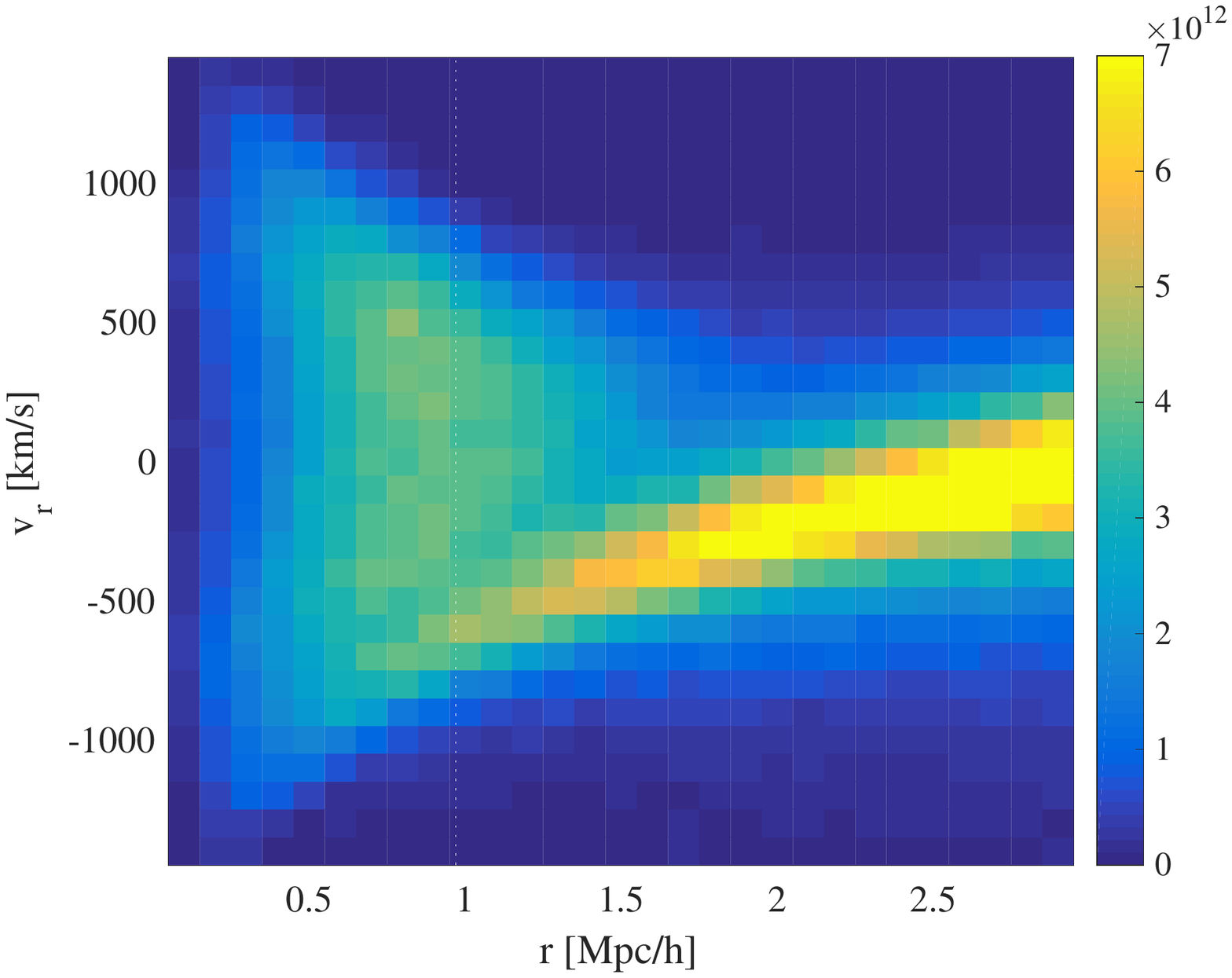}}
\centerline{\includegraphics[width=0.48\textwidth,trim=0 2.5in 0 2.5in]{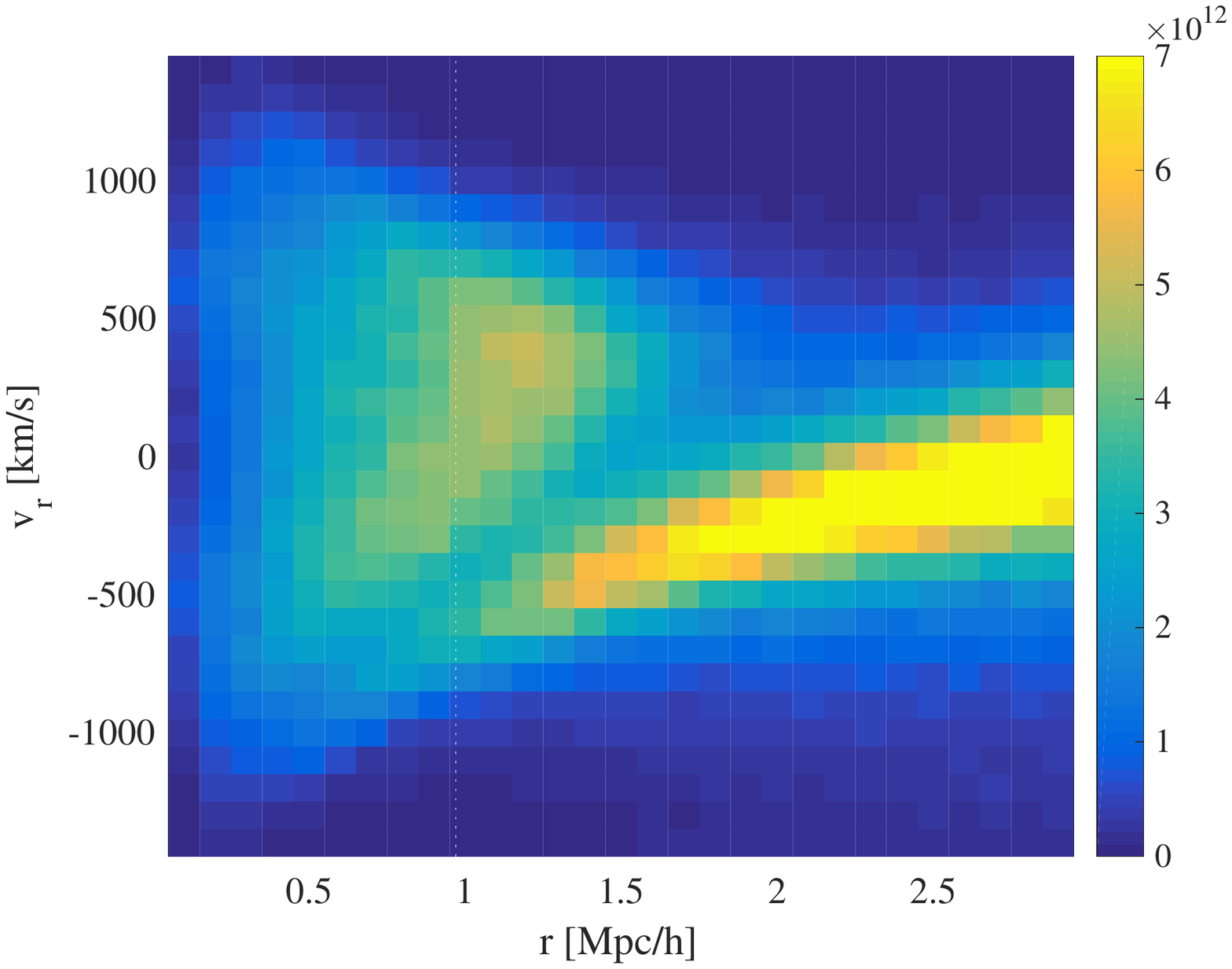}}
\caption{Phase space diagrams around BigMDPL halos with $M_{\rm vir}=1-1.2\times 10^{14} h^{-1} M_\odot$.  The color corresponds to the mass in neighbors (in units of $h^{-1}M_\odot$) at each pixel in the space of $r$ and $v_r$.  We have used a somewhat wider mass bin than in Fig.\ \ref{optmah} in order to improve the statistical uncertainties.  The top panel is for the low $\alpha_{\rm opt}$, corresponding to the blue curve in Fig.\ \ref{optmah}, while the bottom panel is for high $\alpha_{\rm opt}$, corresponding to the red curve in Fig.\ \ref{optmah}.  For this mass range, $r_{\rm vir}$ is depicted by the vertical dotted white line at slightly less than 1 $h^{-1}$ Mpc.  Comparing the two panels, we can see that the sample in the lower panel has a significant amount of bound, virialized mass near splashback located just outside $r_{\rm vir}$. 
\label{phase}}
\end{figure}

If tides are unimportant, then some other explanation is required to account for the slow growth in $M_{\rm vir}$.  To clarify the origin of this behavior, we plot in Fig.\ \ref{phase} the average (stacked) phase space density for the two subsets of high and low $\alpha_{\rm opt}$.   Using the catalog of all Rockstar halos and subhalos with $M_{\rm peak}>5\times 10^{11} h^{-1}M_\odot$, we compute the mass in neighboring objects as a function of  distance and radial velocity relative to each cluster.  For the clusters being considered, with $M_{\rm vir}=1-1.2\times 10^{14} h^{-1} M_\odot$ at $z=0$, the virial radius is approximately $r_{\rm vir}\approx 0.97 h^{-1}$ Mpc.  Fig.\ \ref{phase} immediately explains why the high $\alpha_{\rm opt}$ subset has stopped growing in $M_{\rm vir}$ since $a\sim 0.85$: that subset of clusters has a large portion of splashback mass beyond the nominal $r_{\rm vir}$. Much of that mass just outside $r_{\rm vir}$ was previously inside the virial radius one crossing time in the past, which corresponds to $a\sim 0.85$.  Therefore, mass has indeed been removed from within $r_{\rm vir}$ for these clusters, but not because of tidal stripping, but instead merely because this recently accreted mass is on wide orbits that extend beyond $r_{\rm vir}$. Although we do not have access to the particle data for this simulation, we can estimate the amount of this extra mass using the population of neighboring halos and subhalos as a proxy for dark matter mass.  Very roughly it appears that the splashback mass for the high $\alpha_{\rm opt}$ sample is larger than $M_{\rm vir}$ by about 60\%.  


Therefore, the physical explanation for the slow mass growth in the red curve of Fig.\ \ref{optmah} may actually be quite mundane.  Simply put, these clusters have been assigned the wrong masses.  Their actual physical masses are larger than the quoted virial masses, and therefore it is no surprise that they are more highly biased.  The problem is that the virial mass definition used in most halo finders does not actually measure the bound, virialized mass around a halo (i.e., the mass within the splashback radius), but instead measures the mass within an arbitrarily chosen density threshold.  Relatedly, the quoted masses in the catalog account only for material within a spherical surface, whereas the actual splashback surfaces around simulated halos can deviate significantly from spherical shapes \citep{Mansfield2017,Diemer2017}.  The problem we described above may not be specific to the $\Delta_{\rm vir}$ definition of halo mass, but instead could arise for other similarly arbitrary definitions like 200$c$ or 200$m$.  Indeed, if we repeat the same calculation for halos selected in bins of $M_{200m}$, we again find that the high-$\alpha_{\rm opt}$ sample with high bias has a significant amount of mass located just outside $r_{200m}$. Adopting even lower density thresholds to produce even larger halo radii could suffer the opposite problem of overestimating halo masses, due to uncollapsed mass prematurely being included in the halo, leading to halos with Lagrangian densities well below the spherical collapse threshold. This would similarly generate spurious assembly bias.  
A more physically correct halo mass definition using the splashback feature should avoid such problems and thereby mitigate this spurious behavior in assembly bias.  Fortunately, implementations of splashback halo masses for simulations now exist \citep{Mansfield2017,Diemer2017}, so it should be possible to avoid this problem in future analyses.

This issue with mass definitions may also explain why the assembly bias signal found using mass accretion histories was somewhat weaker than the signal found using halo concentrations, even though mass assembly history and density profile are both related to the same properties of the initial peaks that collapse to form halos.  The splashback radius can be larger or smaller than the arbitrarily chosen overdensity radii like $r_{\rm vir}$ or $r_{200}$ used in halo finders, depending on the physical accretion of mass onto halos, meaning that at all times there are errors in the derived halo boundary and halo mass.  In principle, these errors could possibly generate enough noise in the derived MAH's to erase some of the assembly bias signal that is physically present.

One question that may arise is why the effect of halo mass definitions does not also corrupt the assembly bias signal for higher masses (e.g.\ $M\sim 10^{15} M_\odot$) the way that it does for lower mass clusters.  It is certainly possible to find clusters in this higher mass range whose apparent MAH's exhibit the plateau shown in Fig.\ \ref{optmah}, but their proportion appears to be far smaller among $10^{15} M_\odot$ clusters than it is among $10^{14} M_\odot$ clusters.  We have not explored this question in detail, but a plausible explanation may simply be that clusters with such high mass are much more rare, corresponding to $\sim 3\sigma$ fluctuations of the linear density, rather than $\sim 2\sigma$ fluctuations.  Any cluster with $M_{\rm vir} \approx 10^{15} M_\odot$ that has a significant amount of mass outside $r_{\rm vir}$ would therefore be an even more massive cluster and would correspond to an even rarer fluctuation.  The fraction of such objects therefore should be smaller at $M\sim 10^{15} M_\odot$ than at $10^{14} M_\odot$, simply because the mass function is so much steeper at the higher mass.

In conclusion, high mass halos do indeed exhibit assembly bias as theoretically expected.  Measuring the amount of assembly bias in simulated halos turns out to be affected by the same problem that bedevils attempts to detect assembly bias in real galaxies and clusters: any small errors in determining halo mass can completely overwhelm the intrinsic assembly bias, simply because of the strong dependence of bias on halo mass. In simulations, this challenge may be overcome using physically motivated halo definitions, but it remains to be seen if the observational challenges to detecting this effect in real clusters can be overcome.

\begin{acknowledgments}
ND and MW thank the organizers and participants of the KITP program ``The Galaxy-Halo Connection'' and the NORDITA program ``Advances in Theoretical Cosmology in Light of Data'' for hospitality and for many helpful discussions during the course of this work.  KITP is supported by the National Science Foundation under Grant No.\ NSF PHY17-48958.  
The MultiDark Database used in this paper and the web application providing online access to it were constructed as part of the activities of the German Astrophysical Virtual Observatory as result of a collaboration between the Leibniz-Institute for Astrophysics Potsdam (AIP) and the Spanish MultiDark Consolider Project CSD2009-00064. The MultiDark-Planck (MDPL) and the BigMD simulation suite have been performed in the Supermuc supercomputer at LRZ using time granted by PRACE.
\end{acknowledgments}

\newcommand{\aj}{Astron.\ J.}
\newcommand{\apjl}{\apj\ Lett.}
\newcommand{\jcap}{Journal of Cosmology and Astroparticle Physics}
\newcommand{\mnras}{MNRAS}

%

\end{document}